\documentclass[1p, preprint]{elsarticle}
\usepackage{amsmath}
\usepackage{mathrsfs}
\usepackage{graphicx}
\usepackage{listings}
\usepackage{xcolor}
\usepackage[hidelinks]{hyperref}

\newcommand{\dmavect}[1]{\boldsymbol{#1}}
\newcommand{\dmamat}[1]{\boldsymbol{#1}}
\newcommand{\dmatilde}{\raise.3ex\hbox{$\scriptstyle\sim$}} 

\begin{document}

\date{\today}

\title{A finite element approach to self-consistent field theory calculations of multiblock polymers}

\author[isu]{David M. Ackerman}
\author[ucsb]{Kris Delaney}
\author[ucsb]{Glenn H. Fredrickson}
\author[isu]{Baskar Ganapathysubramanian\corref{cor1}}
\ead{baskarg@iastate.edu}

\cortext[cor1]{Corresponding author}
\address[isu]{Department of Mechanical Engineering, Iowa State University, Ames, Iowa 50011, USA}
\address[ucsb]{Materials Research Laboratory, University of California, Santa Barbara, USA}

\begin{abstract}
Self-consistent field theory (SCFT) has proven to be a powerful tool for modeling equilibrium microstructures of soft materials, particularly for multiblock polymers.
A very successful approach to numerically solving the SCFT set of equations is based on using a spectral approach.
While widely successful, this approach has limitations especially in the context of current technologically relevant applications.
These limitations include non-trivial approaches for modeling complex geometries, difficulties in extending to non-periodic domains, as well as non-trivial extensions for spatial adaptivity.
As a viable alternative to spectral schemes, we develop a finite element formulation of the SCFT paradigm for calculating equilibrium polymer morphologies.
We discuss the formulation and address implementation challenges that ensure accuracy and efficiency.
We explore higher order chain contour steppers that are efficiently implemented with Richardson Extrapolation.
This approach is highly scalable and suitable for systems with arbitrary shapes.
We show spatial and temporal convergence and illustrate scaling on up to 2048 cores.
Finally, we illustrate confinement effects for selected complex geometries.
This has implications for materials design for nanoscale applications where dimensions are such that equilibrium morphologies dramatically differ from the bulk phases.
\end{abstract}


\maketitle

\section{Introduction}
\begin{figure}
  \centering
  \includegraphics{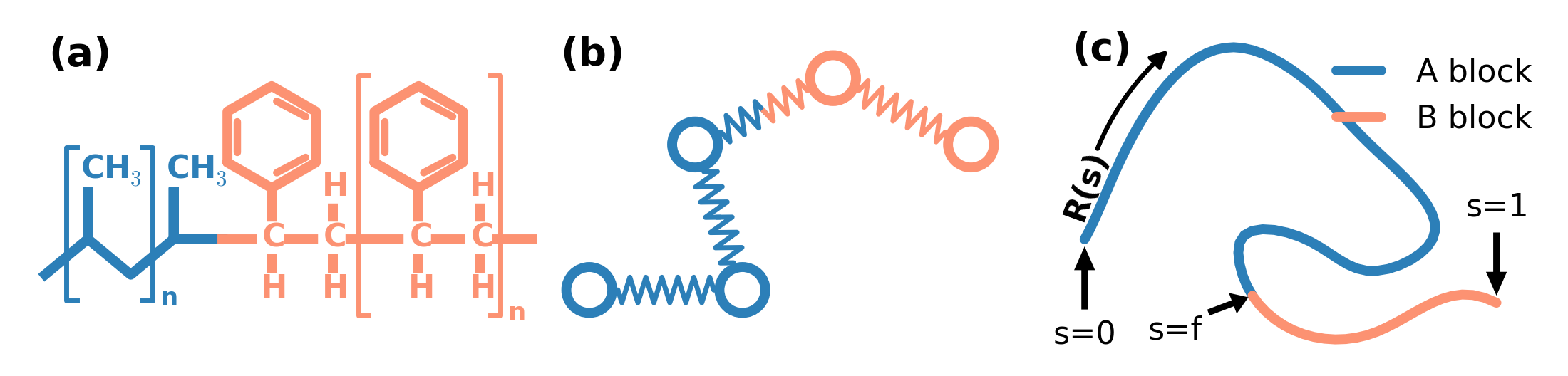}
  \caption{\label{fig:chain_abstraction} Varying levels of abstraction for polymer chain models.
           (a) an atomistic schematic of a diblock system.
           (b) a bead spring model.
           (c) an abstracted, continuous chain model.}
\end{figure}

The morphology of multiblock polymers has been of interest for many years due to potential applications that depend on tailored microstructure.
Numerical simulation can allow study of systems that are outside the limited analytically solvable cases.
However, simulation of equilibrium multi‐block polymer microstructures requires significant computational resources.
A fully atomistic approach treating every atom in the system individually \cite{moule2010, faller2015, bredas2015} is impractical due to the large number of atoms comprising even a single unit cell of a microstructure and prohibitive relaxation times for both bulk materials and non‐periodic, complex geometries.
The computational cost for even a small system is high enough to render this unsuitable as a general tool.
Instead, a coarse-graining approach using an abstracted bead‐spring model (see figure \ref{fig:chain_abstraction}b) as a substitute for the full molecular structure (see figure \ref{fig:chain_abstraction}a) is an often used \cite{doi1986, Faller2011, ganesan2014} alternative.
In this method, each `bead' is actually multiple monomer units with individual beads interacting via carefully designed local and non‐local potentials.
Despite the abstraction, this approach is generally successful at retaining the physics of chain behavior on length scales beyond a nanometer.
While this is far less demanding than a fully atomistic model, it is still a computationally intensive approach for calculating equilibrium microstructures especially for larger and more complex geometries.
A more attractive, continuum approach is the self-consistent field theory (SCFT) method.
SCFT is a mean field theory that starts with the coarse grained chain and interaction models used in particle methods, but transforms the partition function into a field-theoretic framework.

We consider the popularly used model of the continuous Gaussian chain \cite{doi1986} which is well suited for flexible polymers.
This model is based on a linearly elastic chain where the chain stretching is governed by a harmonic potential.
Chain segments interact via pair potentials that are usually assumed to be attractive or repulsive contact interactions (delta functions).
The relevant partition function integral is re-expressed using Hubbard-Statonovich transforms into an integral over auxiliary fields, and is taken to be dominated by a single set of fields (the mean-field approximation).
The procedure is then to solve for these mean fields, which is done iteratively, to obtain the equilibrium field values and with them, the microstructure of the system.
This approach has been used in a wide variety of systems and applications including lithography \cite{Mickiewicz2010}, polymer brushes \cite{milner1988, muller2002, price2012}, self assembly \cite{hur2011, zhu2010}, polymer nano‐composites \cite{Thompson2001}, organic electronics \cite{Jackson2015}, and thin films \cite{Geisinger1999}.
Several recent advances include the use of SCFT in a hybrid SCFT-liquid state theory using charged polymers \cite{Sing2014, Sing2013}.
The addition of electrostatics leads to previously unseen structures with promising potential for energy storage applications due to favorable mechanical and electrical properties \cite{Sing2014}.
SCFT has also been used to study the directed self-assembly approach to lithography \cite{Laachi2015}.
In this work, SCFT simulations were used to identify confining template geometries and polymer formulations that can achieve 10 nm scale patterns targeted by the microelectronics industry, along with acceptable defect levels.
Although much SCFT work has been done using linear AB diblock chains, the method is not limited to those systems.
Multiblock polymers \cite{matsen1999, matsen2000, Guo2008}, star and branched polymers \cite{xu2014, matsen2012}, and tapered diblock polymers \cite{brown2013} have been studied as well.
In the last case, the taper is block of mixed A and B monomers with the ratio changing along the length of the chain.
The addition of the tapered block was found to change the phase behavior of the system, leading to a wider range of stability of the bicontinuous phase.

Spectral methods have been the predominant tool for solving SCFT problems.
The approach is efficient, and has high spatial accuracy. This makes it an excellent choice for many applications.
However, there are applications where the frequency-domain approach (of spectral, and quasi-spectral methods) has limitations; which encourages consideration of alternate real space approaches, like the finite element (FE) method.
First is the ease of handing complex geometries.
While a purely spectral model requires masking techniques for complex geometries, real space methods require no addition actions.
Second, real space methods are not limited to periodic systems and naturally allow the use of heterogeneous and mixed boundary conditions.
Finally, real space methods allow local mesh adaptation to selectively increase the resolution in a targeted position without requiring increased computational effort over the entire system.
This restriction on spectral methods is partially alleviated by use of Chebyshev or other localized bases.
Finite Element approaches, in particular, can incorporate rigorous {\it a posteriori} error estimates (due to the variational treatment) for mesh adaptivity that enable substantial computational gains.
Furthermore, there is a substantial push to design solvers and frameworks (like FASTMath) for real space approaches that are suitable for deployment on next generation exa scale computers.
Motivated by these factors, we develop a real space formulation of the SCFT problem using the finite element method.
The implementation is discussed in detail along with example results and a detailed study of the accuracy of implementation.
Our contribution in this paper include: (a) formulating the SCFT problem in real space using a finite element based variational form, (b) exploring and implementing various high order contour stepping methods, (c) incorporating Richardson extrapolation for multiblock systems, (d) software engineering informed efficient implementation, and (e) illustrative examples (complex geometry, non-periodic domains, scalability studies) of the implementation highlighting the strengths of the method.
The key factors affecting the accuracy of the results are discussed in detail.

The SCFT approach has been covered in great detail elsewhere \cite{fredrickson2013}, but we will review the key points before discussing the finite element implementation.
Broadly, the SCFT approach is a mean field theory where the partition function is dominated by the mean field values of $W(\dmavect {r})$.
The task then becomes finding the value of the mean fields such that:
\begin{equation}\label{eq:scft}
  {\left. \frac{\delta H[W]}{\delta W} \right \rvert}_{W=W^*} = 0
\end{equation}
where H is the Hamiltonian of the system and $W^*$ are the mean field values.
The system considered here is a melt of AB diblock copolymer chains of uniform length.
Extension to multiblock copolymers is straight-forward.
The chains (see figure \ref{fig:chain_abstraction}c) are treated as a continuous space curve with each point along the curve having a contour position s, describing where it is along the length of the chain, and a spatial position $\dmavect r$.
The first part of the chain is a block of type A and the second part is a block of type B.
The crossover point is at s=f ($0<f<1$) with the total length normalized to 1.
Many potentials exist for polymer interactions.
While the treatment is agnostic to the specific form of interaction, we illustrate the framework using the widely used Flory-Huggins model of interaction.
In the Flory-Huggins model, A and B blocks have a repulsive interaction given by $\chi N$, the product of the segmental interaction parameter $\chi$ and the chain length N.
Individual points along the chain are not tracked directly, instead points along the chain are described by a chain propagator $q(s,\dmavect {r})$ which describes the probability of a segment of chain at position s along the contour being located at position $\dmavect{r}$ in space.
The equation for the propagator takes the form of a modified diffusion equation \cite{fredrickson2013} (see Eqn.~\ref{eq:forward_prop}).
Solving this equation yields the propagator values needed to calculate the chain densities and from those, the potential fields.

Section 2 presents a brief overview of the numerical description of the SCFT method.
Following that, section 3 is a detailed treatment of the finite element formulation of the polymer chain propagator.
Section 4 discusses the implementation details.
Results and analysis of the performance are presented in section 5.
We conclude in section 6.

\section{Self Consistent Field Theory Equations}
For the diblock system described above, the Hamiltonian of the system is given by:
\begin{equation}\label{eq:hamiltonian}
\begin{split}
\mathcal{H} = \frac{1}{V} \int d\dmavect r \thickspace
  (
    &\chi N \rho_A(\dmavect r) \rho_B(\dmavect r)
      - W_A(\dmavect r) \rho_A(\dmavect r) \\
    & - W_B(\dmavect r) \rho_B(\dmavect r)
  )
  - ln Q
\end{split}
\end{equation}
where V is the volume of the system, $\rho_A$ and $\rho_B$ are the reduced density fields of the A and B segments, $W_A$ and $W_B$ are the local potential fields for the A and B segments, and Q is the partition function of the system.

Treating the system as incompressible ($\rho_A +\rho_B = 1$) via a Lagrange multiplier gives $H = \mathcal{H} + \int d\dmavect r \lambda (\rho_A +\rho_B - 1)$.
Applying Eqn.~\ref{eq:scft} for the five fields ($\rho_A$, $\rho_B$, $W_A$, $W_B$, and $\lambda$) leads to the well known SCFT equations:
\begin{align}
  W_A = \chi N \rho_B + \lambda \label{eq:SCFTspecieswA} \\
  W_B = \chi N \rho_A + \lambda \label{eq:SCFTspecieswB} \\
  \rho_A + \rho_B = 1 \label{eq:SCFTspeciesLambda} \\
  \rho_A = - \frac{\delta \ln Q}{\delta W_A} \label{eq:SCFTspeciesRhoA} \\
  \rho_B = - \frac{\delta \ln Q}{\delta W_B} \label{eq:SCFTspeciesRhoB}
\end{align}

\begin{figure}
  \centering
  \includegraphics{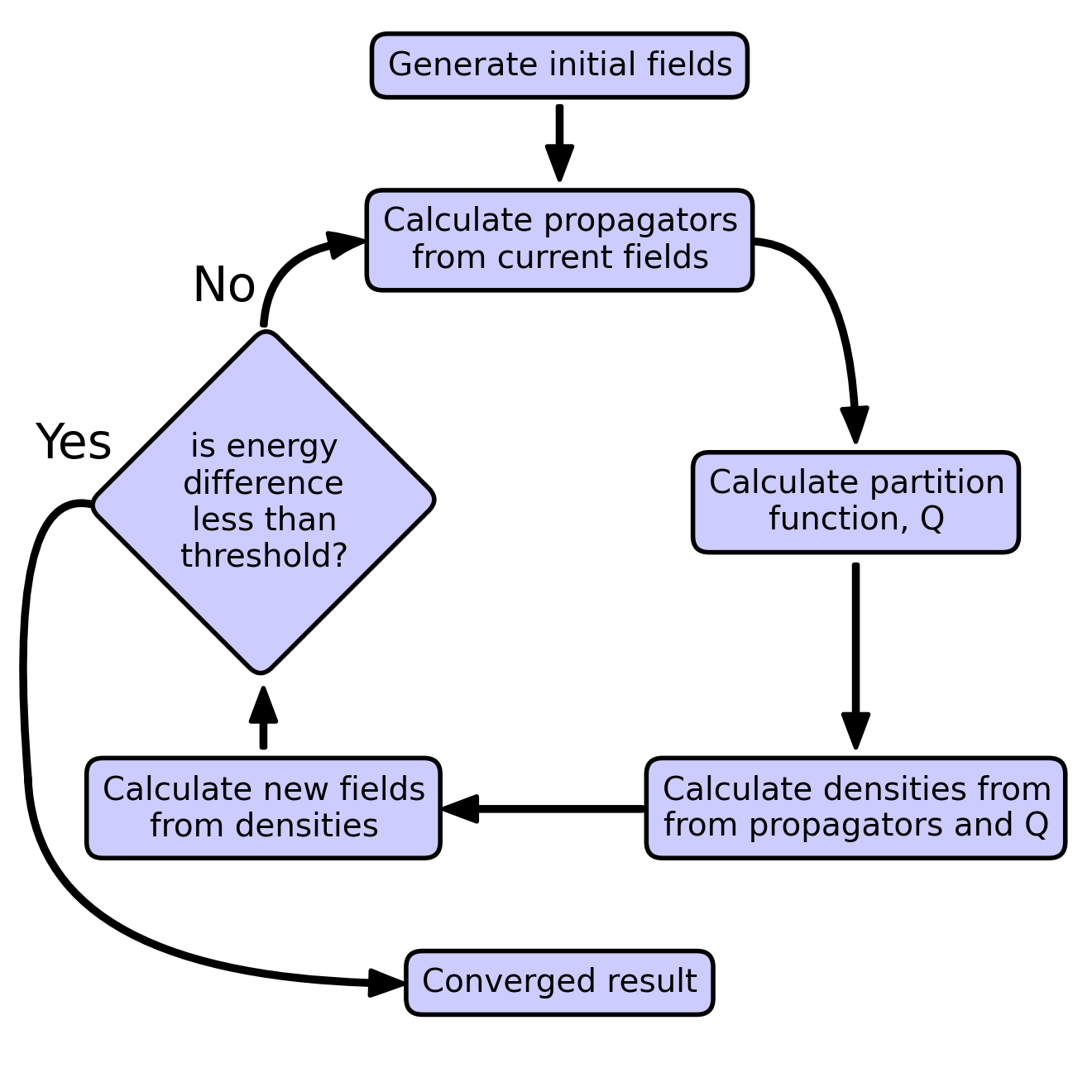}
  \caption{\label{fig:scft_flowchart} Flowchart of SCFT iterative process.}
\end{figure}

Finding the mean field state is done through an iterative process.
The overall sequence of the solution, as illustrated in figure \ref{fig:scft_flowchart}, consists of five main steps: initial field generation, propagator calculation, partition function calculation, density calculation, and field recalculation.
This approach is equivalent to solving a fixed point problem for the field: $W = \mathcal{F}(W)$.

\subsection{Initial Field Generation}
The chain propagator equation is dependent on the potential fields, $W_A$ and $W_B$.
In order to start the process, these must have starting values.
If the starting values are uniform, the gradient term in the modified diffusion equation goes to zero, leaving no driving force for the formation of a microstructure.
To prevent this, there must be some spatial inhomogeneity in the initial values.
Typically this is achieved by choosing random initial values at each point in the system.
The exact values do not matter although a magnitude that is too high will cause instability.
Typically, values in the range [-1,1] are sufficient.
The remaining fields are ($\rho_A$, $\rho_B$, and $\lambda$) are calculated from the potential fields and do not need initial values.

\subsection{Propagator Calculation}
Using the field values, the polymer chains are created as described by a propagator equation.
There are two chain propagators: the forward propagator, $q(\dmavect r, s; [W_A, W_B])$ and the complimentary propagator $q_c(\dmavect r, s; [W_A, W_B])$.
For notational convenience, the dependence on fields $W_A$ and $W_B$ is omitted below.
The forward propagator, $q$, builds the chain from one end, starting at s=0 and moving to s=1:
\begin{equation}\label{eq:forward_prop}
  \frac{\partial}{\partial s}
  q(\dmavect r, s)
  =
  N \frac{{[b(s)]}^2}{6}
  \nabla^2 q(\dmavect r, s) \\
  - W(\dmavect r, s) q(\dmavect r, s)
\end{equation}
where b(s) is the statistical segment length.
Note that s was scaled to a range of [0, 1] and a factor of N is absorbed in W.

In the diblock case the field, $W(\dmavect r, s)$, and the segment length, $b(s)$, are dependent on the position along the chain:
\begin{equation}
 W(\dmavect r,s) \equiv
  \begin{cases}
   W_A,       & 0 \leq s \leq f \\
   W_B,       & f < s \leq 1
  \end{cases}
\end{equation}
\begin{equation}
 b(s) =
  \begin{cases}
   b_A,       & 0 \leq s \leq f \\
   b_B,       & f < s \leq 1
  \end{cases}
\end{equation}
where $b_A$ and $b_B$ are the field and statistical segment values for blocks A and B.

The complimentary propagator, $q_c$, starts from s=1 and moves backwards to s=0:
\begin{equation}\label{eq:backward_prop}
  \frac{\partial}{\partial s}
  q_c(\dmavect r, s)
  =
  \frac{{[b_c(s)]}^2}{6}
  \nabla^2 q_c(\dmavect r, s)
  - W_c(\dmavect r,s) q_c(\dmavect r, s)
\end{equation}
The fields and segment lengths for this propagator are given by:
\begin{equation}
 W_c(\dmavect r, s) \equiv
  \begin{cases}
   W_B,       & 0 \leq s \leq 1-f \\
   W_A,       & 1-f < s \leq 1
  \end{cases}
\end{equation}
\begin{equation}
 b_c(s) \equiv
  \begin{cases}
   b_B,       & 0 \leq s \leq 1-f \\
   b_A,       & 1-f < s \leq 1
  \end{cases}
\end{equation}

For both propagators, the initial condition at s=0 is:
\begin{equation}\label{eq:q0}
  q(\dmavect r, 0)
  =
  q_c(\dmavect r, 0)
  =
  1
\end{equation}

\subsection{Partition Function Calculation}
Once the propagator is computed, a partition function value is calculated:
\begin{equation}
  Q[W_A, W_B]
  = \frac{1}{V} \int d \dmavect r \thickspace q(\dmavect r, 1)
\end{equation}
This can equivalently be defined from the complimentary propagator:
\begin{equation}
  Q[W_A, W_B]
  = \frac{1}{V} \int d \dmavect r \thickspace q_c(\dmavect r, 1)
\end{equation}

\subsection{Density Calculation}
Using this partition function and the propagator values along the chain, the segment densities at position $\dmavect r$ are given by:
\begin{multline}
  \rho_A(\dmavect r, [W_A, W_B])
  =
  - \frac{1}{Q[W_A, W_B]}
  \frac{\delta Q[W_A, W_B]}{\delta W_A(\dmavect r)} \\
  =
  \frac{1}{Q[W_A, W_B]}
  \int_0^{f} ds \thickspace
  q_c(\dmavect r, 1-s) q(\dmavect r, s)
\end{multline}
\begin{multline}
  \rho_B(\dmavect r, [W_A, W_B])
  =
  - \frac{1}{Q[W_A, W_B]}
  \frac{\delta Q[W_A, W_B]} {\delta W_B(\dmavect r)} \\
  =
  \frac{1}{Q[W_A, W_B]}
  \int_{f}^{1} ds \thickspace
  q_c(\dmavect r, 1-s) q(\dmavect r, s)
\end{multline}

\subsection{Field Recalculation}
With the densities at each point known, the new potential fields are calculated using SCFT Eqns.~\ref{eq:SCFTspecieswA} - \ref{eq:SCFTspeciesLambda}.
Solving Eqns.~\ref{eq:SCFTspecieswA} and \ref{eq:SCFTspecieswB} for $\rho_A$ and $\rho_B$ and applying Eqn.~\ref{eq:SCFTspeciesLambda} gives $\lambda$ for each node point (with superscripts denoting which iteration the values are taken from):
\begin{equation}
  \lambda^n = \frac{W_A^{n-1} + W_B^{n-1}}{2} - \frac{\chi N}{2}
\end{equation}
The target fields for iteration n are then
\begin{equation}
  W_A^{n\dagger} = \chi N \rho_B^{n} + \lambda^{n}
\end{equation}
\begin{equation}
  W_B^{n\dagger} = \chi N \rho_A^{n} + \lambda^{n}
\end{equation}
The new field values for this iteration are set using a simple mixing scheme:
\begin{equation}
  W_A^n = W_A^{n-1} + \alpha (W_A^{n\dagger} - W_A^{n-1})
\end{equation}
\begin{equation}
  W_B^n = W_B^{n-1} + \alpha (W_B^{n\dagger} - W_B^{n-1})
\end{equation}
where $\alpha$ is an under relaxation factor ($\alpha \leq 1$).
These values will differ from the previous values if the system is not in a stationary state.
The degree of difference is an indication of convergence to the correct solution.
A threshold value is used to decide when the system is converged.
If the error is less than the threshold the system is considered converged.
If it is greater, the process is started again from step 2 calculating the propagator using the new fields.
This cycle continues until the fields are consistent with the calculated structure.

\section{Propagator formulation}\label{sect:propagator_formulation}
Since the propagator is the most computationally demanding part of the solution process, we discuss the implementation in detail.
This section presents the solution of the propagator using finite elements.
It gives the weak form and matrix form of the propagator equation (Eqn.~\ref{eq:forward_prop}).
Following that, several methods of discretizing the contour derivative are presented using a typical contour stepping solving scheme.

\subsection{Variational Form}
Consider a domain $\Omega$, with boundary $\Gamma_g$ on which Eqn.~\ref{eq:forward_prop} is to be solved.
Let $\mathcal{V} = \mathcal{V}(\Omega)$ denote both the trial and test function spaces.
The variational formulation, i.e. weak form, is stated as following:

Find $q \in \mathcal{V}$ such that for $\forall w \in \mathcal{V}$:  %
\begin{equation}
(w,  \frac{d}{ds} q)  + (\nabla w, G \nabla q) + (w, W(r) q)   =   0
\end{equation}

G is defined from Eqn.~\ref{eq:forward_prop} as $N {[b(s)]}^2 / 6$, and $(.,.)$ is the inner product over the domain $\Omega$.
Defining the bilinear form as $a(u,v) \equiv \int_\Omega \nabla u \nabla v d\Omega $
gives the weak form of the propagator equation:
\begin{equation}\label{eq:prop_weak}
(w,  \frac{d}{ds} q)
 + G a(w, q)
  + (w W(r), q)
  =
  0
\end{equation}

\subsection{Semi-discrete Matrix Form}

We then consider a triangulation, $\mathcal{T}_h$, of the domain $\Omega$.
$\mathcal{T}$ consists of a set of (finite) elements $\Omega_i$, of size $h$, such that $\cup \Omega_i = \Omega$ and $\cap  \Omega_i = \emptyset$.
We consider an approximation of the weak form on the triangulation and approximate space, $\mathcal{V}^h \subset H^h(\Omega)$.
The problem is now to find $q_h \in \mathcal{V}^h$ with the above boundary condition, such that
\begin{equation}\label{eq:prop_galerkin}
  \frac{d}{ds} (w_h, q_h)
  + G a(w_h, q_h)
  + (w_h W_h, q_h)
  =
  0
\end{equation}
for $w_h \in H^h$.

We associate a standard set of basis functions with each element $\Omega_i$.
Thus, $q$ is expanded in terms of the basis functions $\{ N_A\}_{A=1}^{n_b}$, where $n_b$ is the number of basis functions.
The problem now becomes one of finding the nodal values of the unknown quantity, $q$, over the triangulation.
This form is simplified by writing $q$ and $w$ in terms of their nodal values and the basis functions.
Recall that we denote the inner product over the element as $(f,g)$.
The various terms in the Eqn.~\ref{eq:prop_galerkin} can be written in terms of the two matrices $\dmamat{M}_{AB}=(N_A, N_B)$ and $\dmamat{K}_{AB}=G \times a(N_A, N_B)$ and (unknown) vector $\dmavect{q}_{B}=q_B$ to get the semi-discrete matrix form:
\begin{equation}\label{eq:matrix_form}
  \frac{d\dmavect{q}}{ds} \dmamat{M}  + \dmamat{K}\dmavect{q}  + \dmamat{M}W(r)\dmavect{q} = 0
\end{equation}

\subsection{Discrete Form}
The matrix form above is still continuous in the contour variable.
There are numerous possible ways to discretize the contour.
The primary difference between methods is the order of accuracy.
A higher order of accuracy is desired because it allows the use of large $\Delta s$ corresponding to fewer evaluations of $q$ for a given length of the chain.
Since the propagator requires the major computational effort, reducing the number of propagator evaluations is a highly effective way to improve performance.
Mulitple numeric schemes have been proposed\cite{rasmussen2002, cochran2006, ranjan2008, tzeremes2002} and there have been several comparisons for pseudospectral algorithms\cite{Stasiak2011, audus2013}.
To study the finite element method we choose five different discretization schemes: Backward Euler (BE); Crank-Nicolson (CN); and the backward differentiation formulas 2, 3, and 4 (BDF2, BDF3, and BDF4).
All methods require splitting the equation into a finite number of steps with a spacing of $\Delta s$.
The BE is 1st order accurate in $s$, CN and BDF2 are 2nd order accurate, BDF3 is 3rd order, and BDF4 is 4th order.

\subsubsection{Backward Euler}
The backward Euler approximation for a function of the form $\partial q/\partial s = f$ is:
\begin{gather}\label{backwardEuler}
  \frac{q^{n} - q^{n-1}}{\Delta s} \approx f^{n}
\end{gather}
where the superscript denotes the contour step.

This is the simplest discretization and applying it to Eqn.~\ref{eq:matrix_form} gives:
\begin{equation}\label{prop_be}
  \dmamat M  \dmavect q^{n}
  + \Delta s \dmamat K \dmavect q^{n}
  + \Delta s \dmamat M W(r) \dmavect q^{n}
  = \dmamat M  \dmavect q^{n-1}
\end{equation}
for n from 1 to the number of contour points.

This equation is solved first by using the initial condition (Eqn.~\ref{eq:q0}) and the field values to solve for the values of $q^1$ across the entire system.
This value is then used to find $q^2$, and so on until the propagator values for the entire chain have been calculated.

\subsubsection{Crank-Nicolson}
The Crank-Nicolson approximation for a function of the form $\partial q/\partial s = f$ is:
\begin{gather}\label{Crank-Nicolson}
  \frac{q^{n} - q^{n-1}}{\Delta s}
  \approx
  \frac{1}{2} (f^{n} + f^{n-1})
\end{gather}
where the superscript denotes the contour step.

Following the same procedure given for the BE case, the matrix form is

\begin{align}
  \dmamat M  \dmavect q^{n}
  + & \frac{\Delta s}{2} \dmamat K \dmavect q^{n}
  + \dmamat M \Delta s W(r) \dmavect q^{n}
  = \nonumber \\
  & \dmamat M  \dmavect q^{n-1}
  - \frac{\Delta s}{2} \dmamat K \dmavect q^{n-1}
  - \dmamat M \Delta s W(r) \dmavect q^{n-1}
\end{align}
for n from 1 to the number of contour points.
As before the initial condition is $\dmavect q^0=1$ and the values along the chain are solved sequentially.

\subsubsection{Backward Difference 2, 3, and 4}
The backward difference formulas 2, 3, and 4 all follow a similar format.
The number indicates both the order of accuracy in contour step and the number of previous contour terms required for the formulation.
The need for more than one previous contour point creates a complication not present with the BE and CN methods.
Points at the beginning of the discretized chain may not have a sufficient number of previous points to utilize these methods.
This `startup problem' is not unique to the BDF methods; It applies to all schemes requiring more than one previous point.
A convenient approach to solving this is the use of Richardson Extrapolation \cite{Richardson1911, ranjan2008}.
Application of Richardson Extrapolation for BDF2, BDF3, and BDF4 is given in detail in the appendix.
The BDF equations are given below:

The equation for BDF2 is:
\begin{equation}
  \dmamat M \dmavect q^{n}
  + \frac{2}{3} \Delta s \dmamat K \dmavect q^{n}
  + \frac{2}{3} \Delta s \dmamat M W(r) \dmavect q^{n}
  =
  \frac{4}{3} \dmamat M \dmavect q^{n-1}
  - \frac{1}{3} \dmamat M \dmavect q^{n-2}
\end{equation}

The equation for BDF3 is:
\begin{equation}
\dmamat M \dmavect q^{n}
  + \frac{6}{11} \Delta s \dmamat K \dmavect q^{n}
  + \frac{6}{11} \Delta s \dmamat M W(r) \dmavect q^{n}
  =
  \frac{18}{11} \dmamat M \dmavect q^{n-1}
  - \frac{9}{11} \dmamat M \dmavect q^{n-2}
  + \frac{2}{11} \dmamat M \dmavect q^{n-3}
\end{equation}

The equation for BDF4\cite{cochran2006} is:
\begin{equation}
  \dmamat M \dmavect q^{n}
  + \frac{12}{25} \Delta s \dmamat K \dmavect q^{n}
  + \frac{12}{25} \dmamat M \Delta s W(r) \dmavect q^{n}
  =
  \frac{48}{25} \dmamat M \dmavect q^{n-1}
  - \frac{36}{25} \dmamat M \dmavect q^{n-2}
  + \frac{16}{25} \dmamat M \dmavect q^{n-3}
  - \frac{3}{25} \dmamat M \dmavect q^{n-4}
\end{equation}

\section{Finite Element Implementation}
The finite element approach is implemented with a focus on the primary advantages of this method over the spectral approach.
The primary goal is an efficient solution of large problems and those with irregular shapes or within confinement.
Support for arbitrary shapes is inherent in the finite element method.
Large problems are efficiently handled through a highly parallel implementation.
To enable solving of large and complex domains, we use a parallel in-house FEM library built upon the PETSc library \cite{petsc-web-page,petsc-user-ref,petsc-efficient}.
The library handles the finite element backend details and the application code implements the actual SCFT science.
PETSc routines are used for the solving of the system described above.
By default, solving is done using the generalized minimal residual method.
Arbitrary meshes are supported and parallel domain decomposition of meshes is performed using ParMETIS \cite{parmetis-web-page}.
This implementation supports multiple options for boundary conditions.
Periodic, Neumann, and Dirichlet conditions are all easily supported through the FEM library.
Both uniform and mixed, non-homogeneous boundary conditions can be applied.

\subsection{Spatial Discretization}
The accuracy of the solution has two primary components - a component from the accuracy of the contour discretization of the chain, and a component from the accuracy of the spatial discretization.
The choice of the number of contour points and the selection of a contour discretization scheme allow control of the accuracy in contour.
These considerations are similar for both the spectral and real space approaches.
A primary disadvantage of the real space approach is that it lacks spectral accuracy in space.
Spatial accuracy in the finite element method is controlled by both the number of spatial elements and the order of the basis functions used.
The number of elements used can be increased, typically without significant difficulty regardless of the choice of structure.
An alternate approach of increasing the basis function order utilizes the same formulation given above.
\footnote{The library handles the details of the basis function internally, allowing a basis order agnostic code to be used for calculation of the $\dmamat{M}$ and $\dmamat{K}$ matrices.}

Under reasonable bounded and smoothness assumptions on $W$, one can prove convergence estimates for the solution in terms of the triangulation (element length, $h$), the order of the basis function $\beta$ and the order of the contour stepper used $\alpha$ as
\begin{equation}
  \lVert u_{\Delta s}^h - u_e \rVert \le
  c_1 {\Delta s}^{\alpha} + c_2 h^{\beta}
\end{equation}
where $u_{\Delta s}^h$ is the computed solution, $u_e$ is the true solution, and $c_1$ and $c_2$ are constants.

\section{Results}
We first investigate the accuracy of the model with variations in spatial and contour discretization.
We also show several results on non-uniform meshes, which illustrates one of the strengths of this formulation.

\subsection{Comparison of Discretizations}
\begin{figure}
  \centering
  \includegraphics{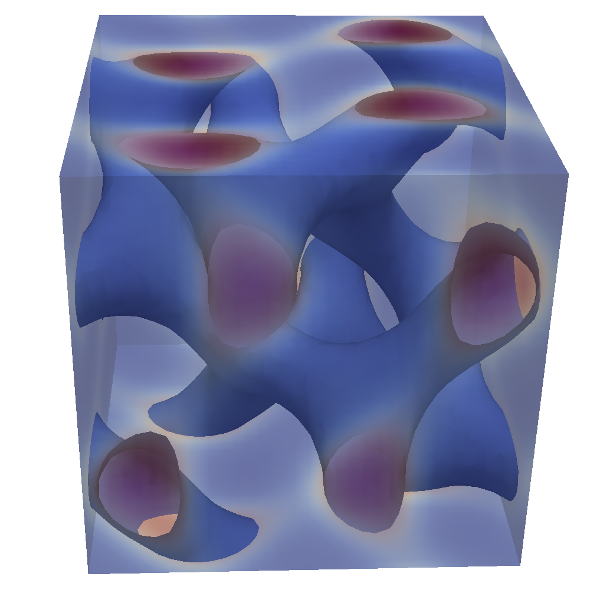}
  \caption{\label{fig:gyroid} Gyroid phase of diblock polymer. See text for parameters.
           }
\end{figure}

To better understand the discretizations described above, we explore the discretization in both space and contour.
In order to compare values, we need a metric for accuracy that can be readily compared across different calculations.
We use the value of the partition function, Q, as the metric of accuracy of the solver.
Since it is an integration of the end result of the propagator solve, it is a measure of the entire finite element solution process.
As a basis for comparison, we use a cubic system with an edge length of 8.82 times the polymer radius of gyration ($R_g$).
The diblock chain is 40\% A and 60\% B (which corresponds to an A block fraction, f, of 0.4) with an interaction parameter of $\chi N$ = 14.4.
This corresponds to a 3D gyroid phase of the diblock polymer (see figure \ref{fig:gyroid}).
In this case, $b_A = b_B$, making the A and B blocks indistinguishable.
The spectrally determined Q value is 5.32583.

\begin{figure}
  \centering
  \includegraphics{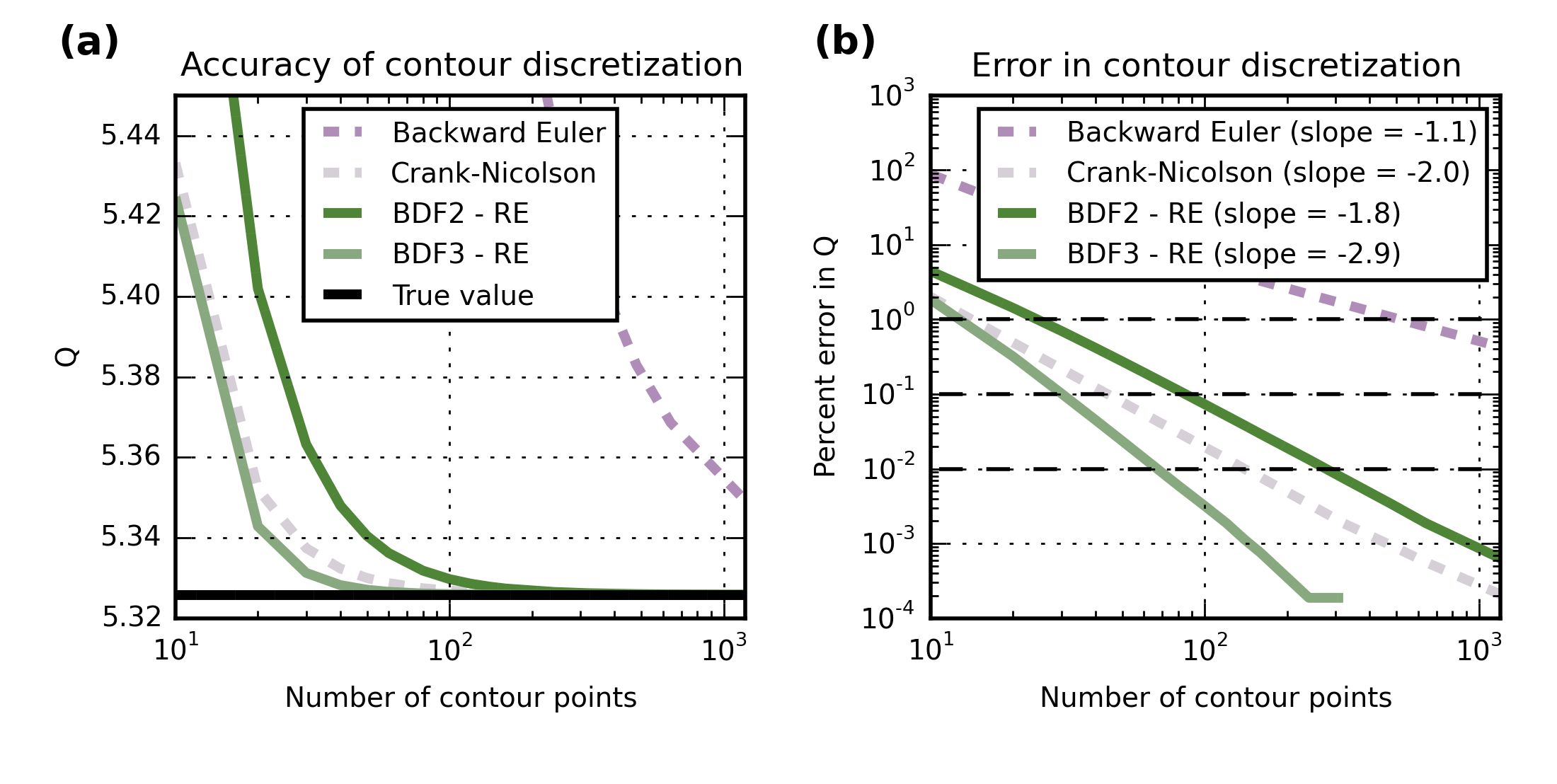}
  \caption{\label{fig:q_convergence} Convergence of Q with contour point counts.
           (a) shows the actual Q results with increasing number of contour points.
           The values eventually converge to the same result as the other methods.
           (b) shows the error in the Q value with varying number of contour points.
           The three dashed lines indicate error values selected for comparison in Table \ref{tab:q_error}.
           }
\end{figure}

First, we look at the contour discretization.
The computational time required for a solution scales linearly with the number of contour points.
Thus the goal is to have the fewest number of contour points necessary for the desired accuracy.
Multiple discretization methods were given in section \ref{sect:propagator_formulation}.
The primary difference among these methods is the order of accuracy.
Figure \ref{fig:q_convergence}(a) shows the convergence of Q for the five discretization methods discussed above.
As expected, the higher order methods converge faster to the desired value.
The simplest backward Euler(BE) approach shows very poor convergence and is mostly unsuitable for use, while the BDF3 method rapidly converges.
Since the propagator solving is the rate limiting step in this problem, it is worth the extra complexity of a higher method in order to reduce the number of contour points.
Figure \ref{fig:q_convergence}(b) shows the percent error for each method at selected numbers of contour points.
The convergence of different methods is readily seen.
Arbitrarily selected accuracies of 1\%, 0.1\%, and 0.01\% are noted on the plot and shown in table \ref{tab:q_error}.
For the coarsest tolerance of 1\%, the BE method requires 540 contour points, while the BDF2 method requires only 26.
If we take the cost of calculation per contour point to be the same across each method (it is within ~15\%), the BDF2 calculation will be over 20 times faster than the BE method in achieving a result within 1\% accuracy.
The BDF3 method is twice as fast as the BDF2 method in achieving 1\% error.
The difference is even more pronounced at the 0.01\% accuracy, where the BDF3 method is four times faster than the BDF2 method.

\begin{figure}
  \centering
  \includegraphics{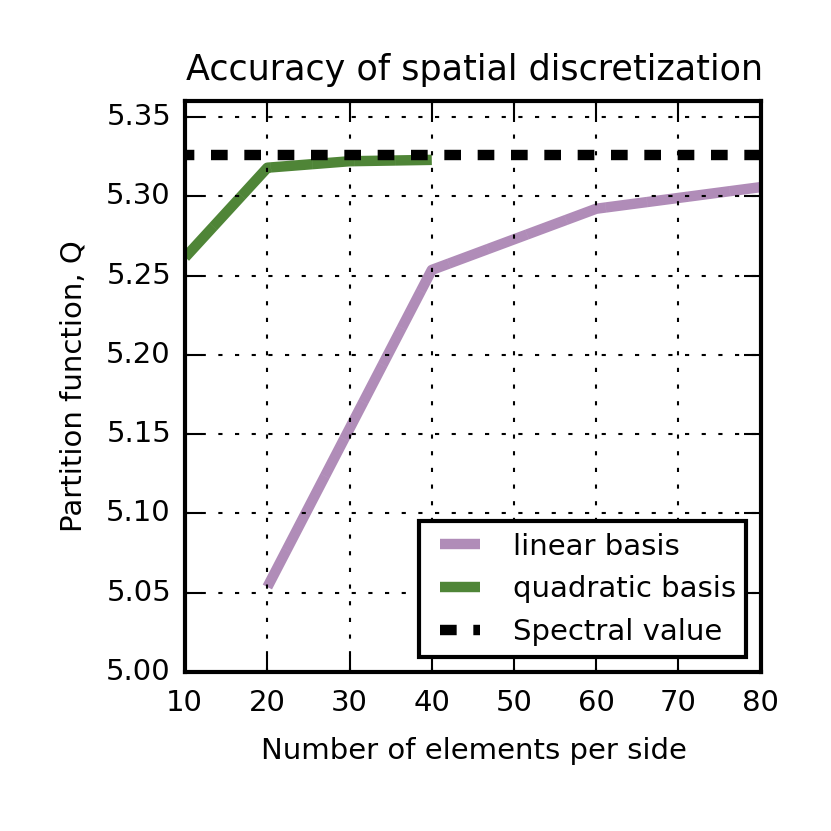}
  \caption{\label{fig:spatial_convergence} Spatial convergence of Q values.
           Calculations done on gyroid phase using 150 contour points and BDF3 contour stepper with Richardson extrapolation.
           The number of elements is the number per side of the cubic system.
           }
\end{figure}

Second, the spatial discretization is addressed.
As mentioned previously, the lack of spectral accuracy makes the order of the basis function important for the accuracy of a give spatial discretization.
Both first and second order basis functions were used.
Higher order, and even spectral basis functions can potentially be used.
Figure \ref{fig:spatial_convergence} shows the resulting Q values for linear and quadratic basis functions using the BDF3 contour steppers.
These calculations were done for a range of cubic systems with the given number of elements.
For all calculations, 150 contour points were used.
Based on the results of the contour testing, this is sufficient to lead to a converged state, so any error in these results is due to the spatial discretization.
It can be clearly seen that the quadratic basis functions are more accurate for a given number of elements and the results converge faster with additional elements, as expected.

\begin{table}[]
\centering
  \caption{\label{tab:q_error} Number of contour points required for a given accuracy.
           All calculations are for a gyroid phase with $64^3$ nodal points, f=0.4, and quadratic basis functions.
           Values are interpolated from data in figure \ref{fig:q_convergence}(b).
           Note that the Backward Euler method did not reach 0.01\% error in less than 10,000 points.
           }
  \begin{tabular}{l|l|l|l}
                                         & 1\%    & 0.1\%  & 0.01\%   \\
                                         & error  & error  & error   \\
    \hline Backward Euler                & 540        & 5114         & --              \\
    \hline Crank-Nicolson                & 15         & 46           & 143              \\
    \hline Backward difference formula 2 & 26         & 85           & 286              \\
    \hline Backward difference formula 3 & 13         & 32           & 69
\end{tabular}
\end{table}

\subsection{Arbitrary Domain Shapes}
\begin{figure}
  \centering
  \includegraphics[width=\textwidth]{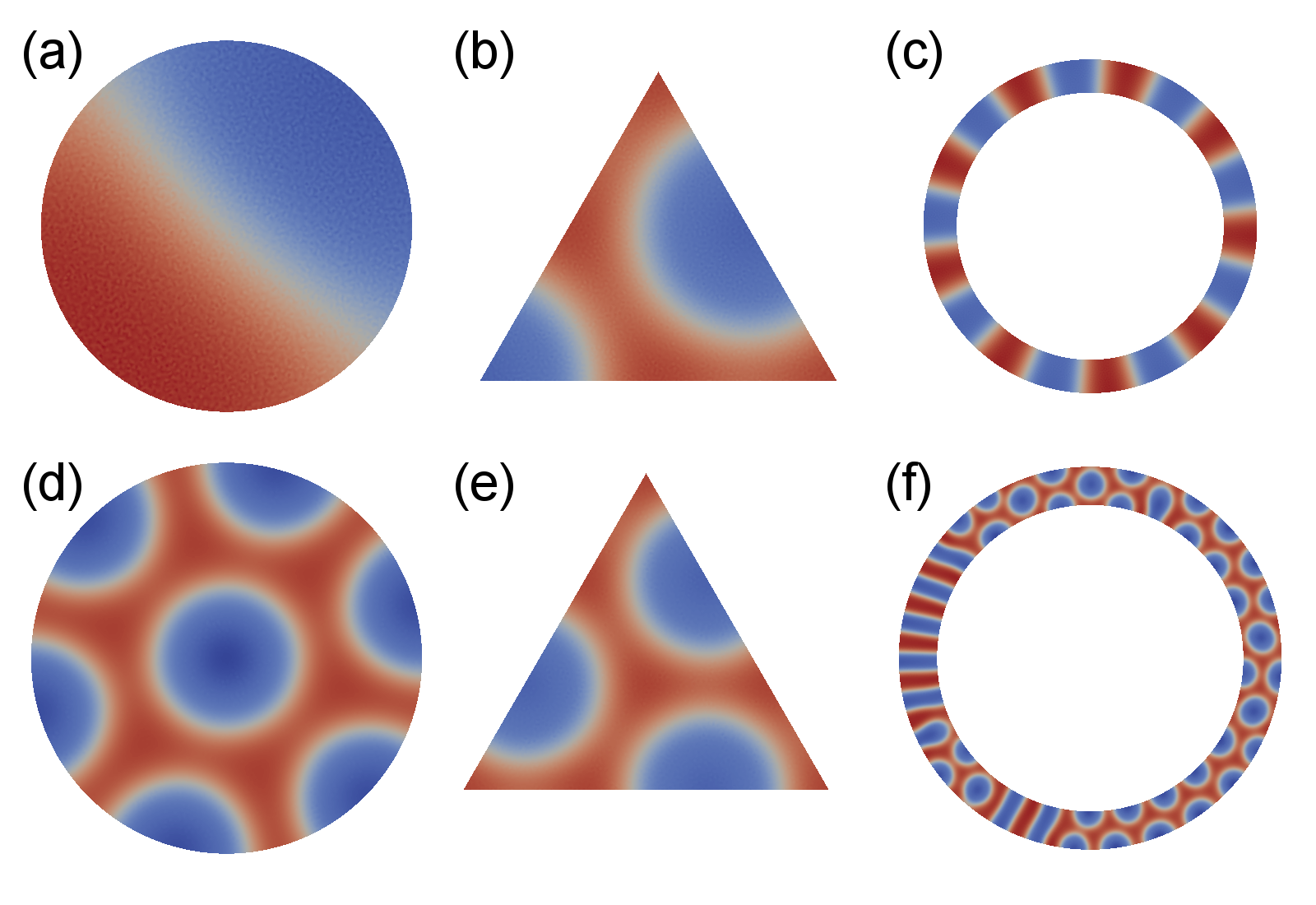}
  \caption{\label{fig:example_structures_2d_1}2D structures using a non-uniform, non-square mesh.
           For all cases $\chi N$ = 14.4, f = 0.4, and zero-flux boundary conditions were applied.
           The circle in (a) has a radius of 1$R_g$, while (d) has a radius of 4$R_g$.
           The triangle in (b) has an edge length of 4$R_g$, while (e) has an edge length of 6$R_g$.
           The annulus in (c) has an outer radius of 5$R_g$ and an inner radius of 4$R_g$, while (f) has an outer radius of 18$R_g$ and an inner radius of 14.4$R_g$.
           }
\end{figure}

\begin{figure}
  \centering
  \includegraphics[width=\textwidth]{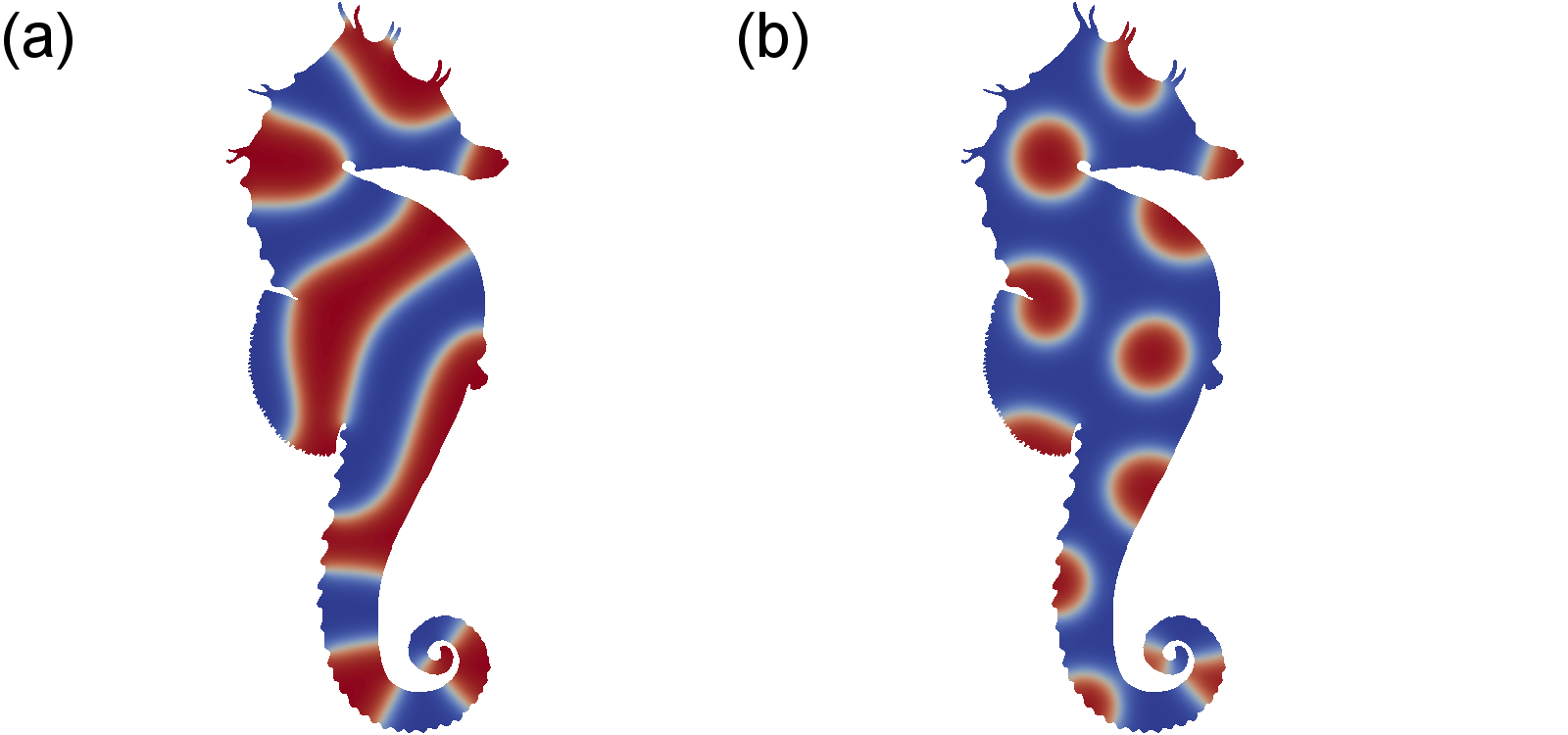}
  \caption{\label{fig:example_structures_2d_3}Arbitrarily shaped 2D structures.
           For both cases $\chi N$ = 25 and the height is approximately 20 $R_g$.
           In (a), f = 0.5 and in (b), f = 0.3.
           Zero-flux boundary conditions were applied.
           The effect of the boundary conditions can be seen in the curved distortion of the A/B interface near the surface.
           }
\end{figure}

\begin{figure}
  \centering
  \includegraphics[width=\textwidth]{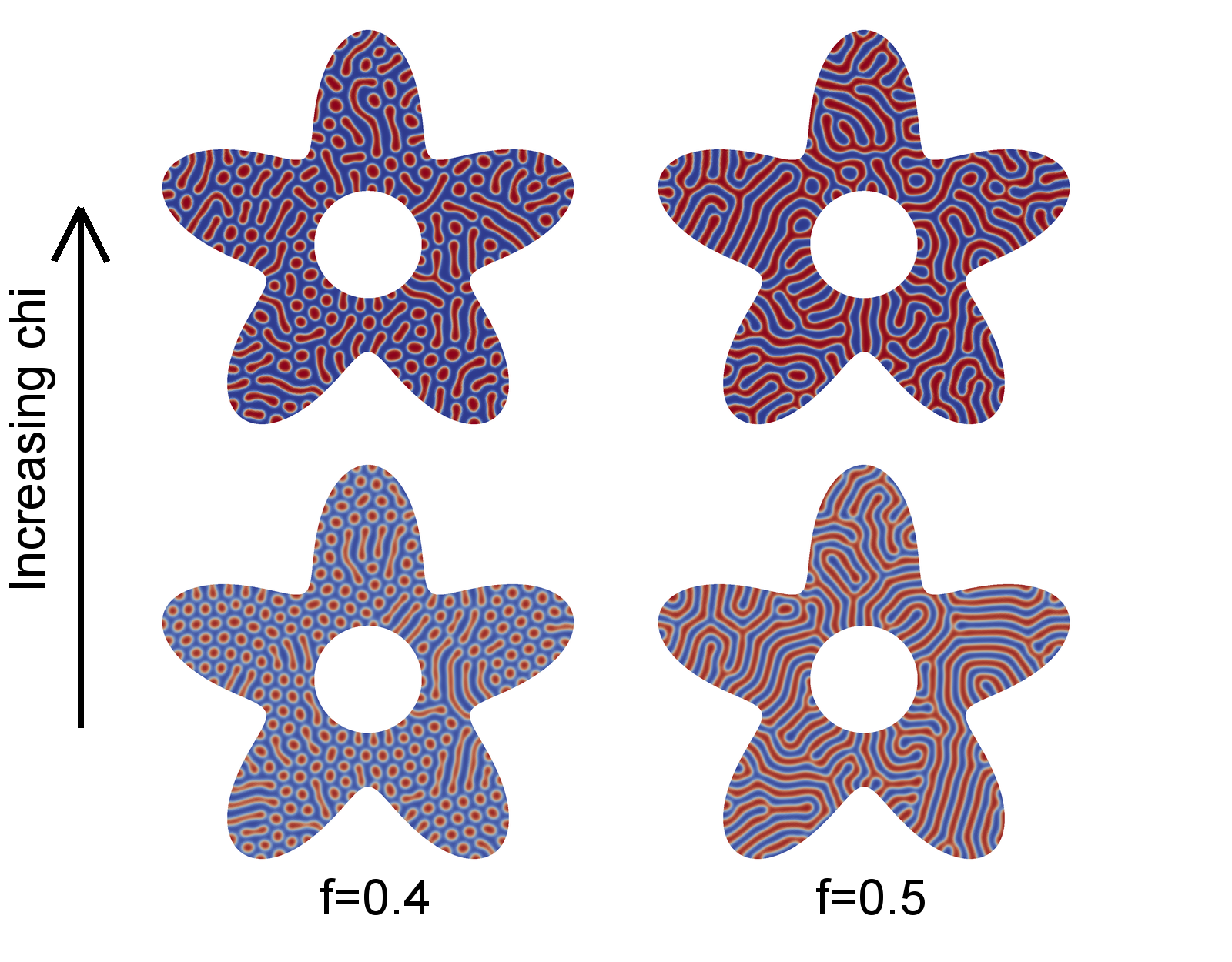}
  \caption{\label{fig:example_structures_2d_2}2D structures using a non-uniform mesh to show variation with fraction of A chain (f) and $\chi N$ values.
          The top row of images has $\chi N$ = 25 while the bottom row has $\chi N$ = 14.4.
          In all cases, the distance from the center to the outer most point is 10 $R_g$ and zero-flux boundary conditions were used.
          }
\end{figure}

\begin{figure}
  \centering
  \includegraphics[width=\textwidth]{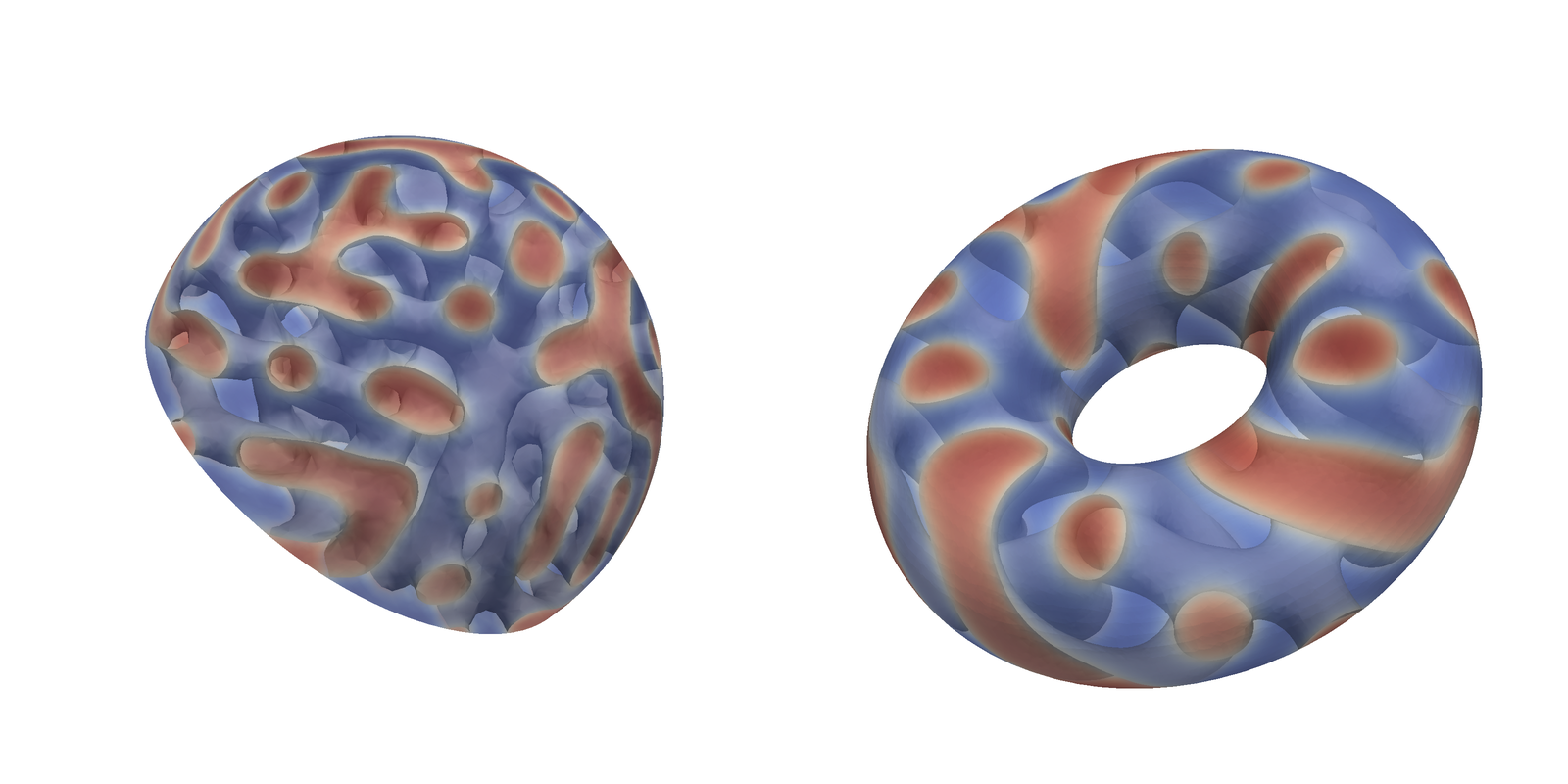}
  \caption{\label{fig:example_structures_3d}3D structures of sphere and torus.
          In each case f = 0.4 and $\chi N$ = 14.4, conditions corresponding to gyroid phase stability in the bulk.
          Zero-flux boundary conditions were applied.
          }
\end{figure}

A key advantage of the finite element implementation is the ability to model arbitrary geometries with no changes.
This allows the calculation of structure on physically meaningful domains rather than just a bulk structure.
Applications of non-bulk shapes include thin fibers where the cross section is small enough that the bulk phase does not form, nanoparticles with dimensions below the bulk lattice spacing, and the previously mentioned directed self-assembly for lithography.
It also allows investigation of the effects of confinement on the structures adopted by the chains.
Figures \ref{fig:example_structures_2d_1} - \ref{fig:example_structures_3d} shows results of several shapes in both 2D and 3D (parameters are listed in the caption).
In the images, the blue regions are areas of high A block concentration and the red regions are areas of high B block concentration.
The top row of images differs from the bottom by the size of the mesh.
As can be seen in the annulus case, the structure adopted can be dependent on the system size due to confinement effects.


\subsection{Scaling with Number of Processors}
Another key advantage of the finite element approach is the ability to run very large problems on multiple processors with high efficiency.
For most periodic systems or unit cell calculations, this is not an important consideration.
However, for complex geometries or more complicated chain models, efficient scaling allows modeling large systems.
The benefit comes from the near linear scaling of the finite element approach.
This allows running problems on a wide range of system sizes.

\begin{figure}
  \centering
  \includegraphics{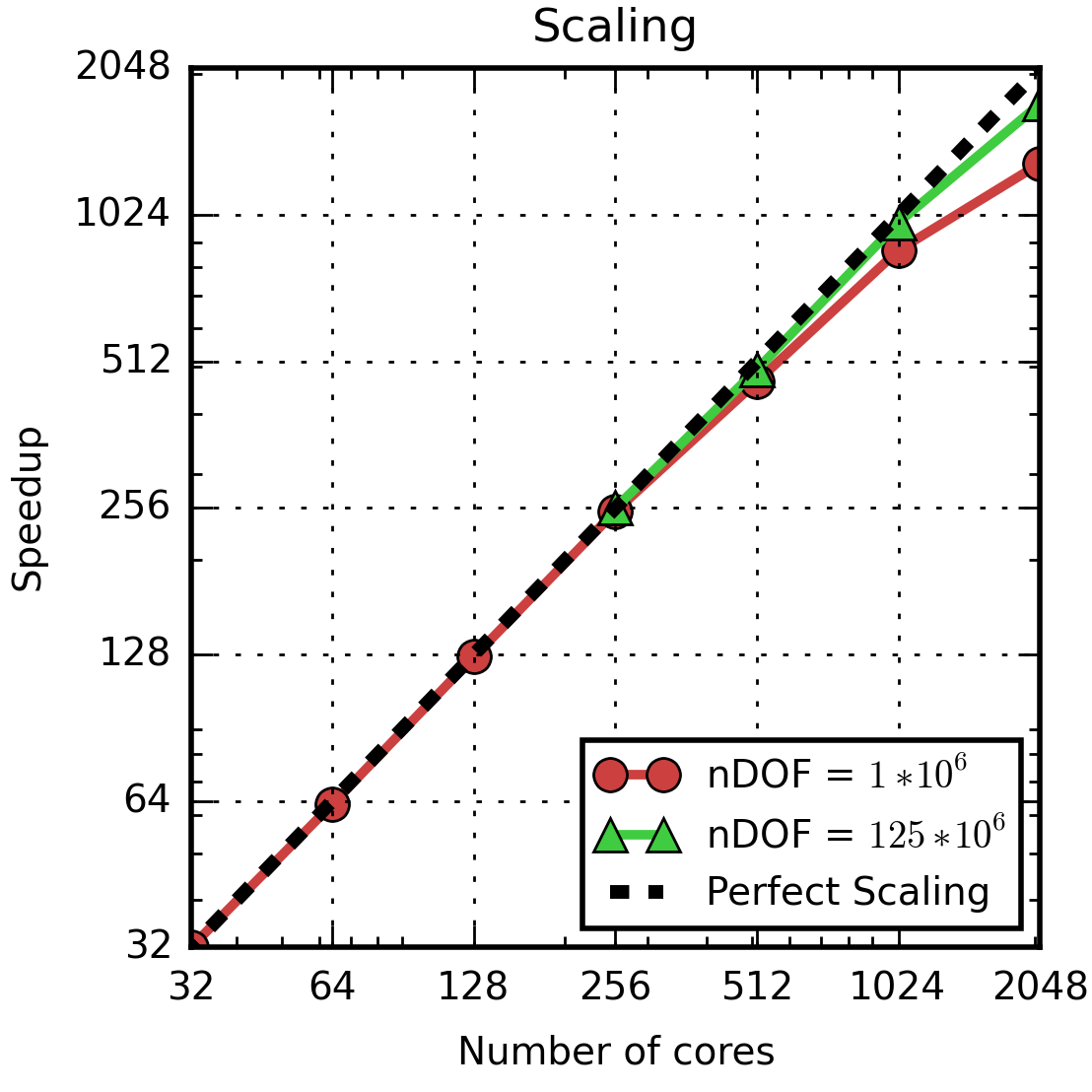}
  \caption{\label{fig:scaling} Strong scaling results using quadratic basis functions on the blue waters system.
           nDOF = number of degrees of freedom being solved in the system.
           Dashed black line shows perfect linear scaling.}
\end{figure}

Figure \ref{fig:scaling} shows scaling results up to 2048 cores using this model on a diblock polymer system.
These results were generated on the Blue Waters system \cite{blue_waters1}.

\section{Conclusion}
The finite element method discussed in this paper is an alternative to the standard spectral and pseudo-spectral methods for self-consistent field theory calculations.
The use of this alternate approach has advantages when considering large problems with non-periodic geometries.
We have discussed details of the implementation relevant to the accurate simulation of block copolymer systems.
Use of a finite element method for SCFT allows easy calculation of self-assembled structures in complex geometries.
This has implications for the study of thin film and other confined systems beyond a simple bulk melt.
The scalability of the approach allows larger structures to be calculated, enabling simulation of physically relevant sizes.
While our focus is on the common diblock case, the method can be applied to multiblock polymers with an arbitrary number of blocks and non-linear architectures.
Beyond the benefits shown here, the SCFT approach also has potential application to chain models other than the Gaussian chain, which are becoming of more interest as the power of computational resources expand.

\section*{Acknowledgments}
DMA and BG were partially supported by NSF 1435587.
KTD and GHF were partially supported by the National Science Foundation grants DMR-1332842 and DMR-1506008.

\section{Appendix: Richardson Extrapolation}
\begin{figure}
  \centering
  \includegraphics{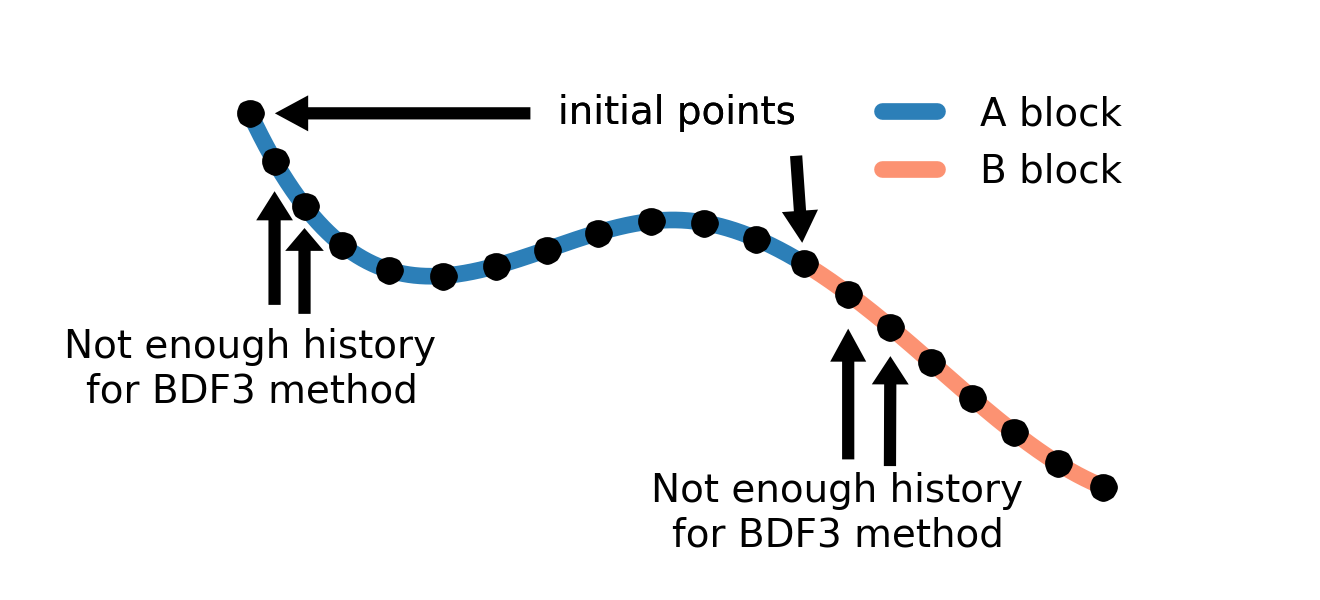}
  \caption{\label{fig:startup_failure}Diagram of diblock copolymer indicating start up problem for BDF3 contour stepping method.}
\end{figure}

The contour discretization described in \ref{sect:propagator_formulation} is used to discretize the propagator equation in the contour variable.
Higher order schemes are desirable because they yield greater accuracy for a given number of discrete points.
As mentioned previously, a problem arises with higher order discretization schemes which require more that a single previous contour point.
At the start of the chain, only one point - the initial condition - is available.
Any method that requires more than one previous point will not have enough previous points to use in the calculation of the early contour points.
Similarly, at the second point, only two points are available - the initial condition and the first point.
For the propagator equation of diblock chains, this startup problem is also an issue in the switch from one block to another.
In this case, there may be enough previous points to use the discretization scheme, however the previous points were created under the effect of different fields so it is physically incorrect to use them as part of the new block.
Fig. \ref{fig:startup_failure} illustrates this problem schematically for the Backward Difference 3 (BDF3) method, which is 3rd order accurate but requires three previous contour points.

The most obvious solution to this startup problem is to simply use a lower order scheme for the first point(s) where there are insufficient points for the higher order method.
In Fig. \ref{fig:startup_failure}, for example, it would be possible to calculate point 2 using a 1st order accurate Backward Difference 1 (BDF1) method and then calculate point 3 using a 2nd order accurate Backward Difference 2 (BDF2) method.
All remaining points would be calculated using the 3rd order BDF3 method.
This is appealing in its simplicity, however it leads to poor results as the errors in the beginning of the chain propagate along the entire length.
In effect, the use of higher order methods later in the chain only serves to preserve the error created in the first point.
An alternative method is the use of Richardson Extrapolation (RE)\cite{Richardson1911, ranjan2008} for the early points.
This method uses solutions from smaller contour steps to build up the accuracy of the early points to match the higher order desired.
Once the early points have been calculated using the RE method, later points are computed using the higher order method as normal.
Through use of this method, it is possible to preserve the higher order accuracy.
Although this method requires greater computational expense for the first points, it can still result in lower overall cost for a given accuracy due to the lower number of points required by the higher order methods.

Briefly, the Richardson method requires the solving for a point with a smaller contour step and the merging the values to create a higher order result.
For the propagator, treat $q^{*}(s)$ as the true value at contour point `s', and $q(s;\Delta s)$ as the value at contour point `s' obtained using a discrete contour step of $\Delta s$.
An Nth order accurate discretization is given by:

\begin{equation}
  q(s;\Delta s) = q^{*}(s) + c (\Delta s)^n + O((\Delta s)^{n+1})
\end{equation}
with c as a constant and $O((\Delta s)^{n+1})$ denoting terms of order N+1 in $\Delta s$.
If we calculate a new value $q(s; h \Delta s)$ with $h<1$ we will get a value calculated with more contour points and thus more accuracy.
Subtracting the results of the two equations to define a new value, we find that the Nth order error term has canceled out, leading to a value that that is accurate in order N+1:
\begin{align}
  q^{RE} & = h^n q(s; \Delta s) - q(s; h \Delta s) \nonumber \\
         & = h^n \left ( q^{*}(s) + c (\Delta s)^n + O((\Delta s)^{n+1}) \right ) \nonumber \\
         & \quad  - \left (q^{*}(s) + c (h \Delta s)^n + O((\Delta s)^{n+1}) \right ) \nonumber \\
         & = \left ( h^n - 1 \right ) q^{*}(s) + O((\Delta s)^{n+1})
\end{align}
Rearranging this gives an expression with the order N term removed:
\begin{equation}\label{eq:re}
  \frac{h^n q(s;\Delta s) - q(s;h \Delta s)}{h^n - 1}
  = q^{*}(s) + O((\Delta s)^{n+1})
\end{equation}
This process can be repeated in stages to remove the N+1, N+2, ... order terms.
Below we show the application of RE to the backward difference 2, 3, and 4 schemes as well as results with and without RE to illustrate the improvement.

\subsection{Backward Difference 2}
The BDF2 scheme is 2nd order accurate in contour and requires two previous contour points.
In order to use BDF2 with a contour step of $\Delta s$ for the calculation of the propagator, we need to calculate the first point to 2nd order accuracy
It should be noted that the second order accuracy can be achieved with a single point using the previously described Crank-Nicolson scheme.
The use of RE can be avoided in the BDF2 scheme by simply calculating the first point using Crank-Nicolson.
Nevertheless, since it is the simplest example, we present the BDF2 scheme with the extrapolation performed using the first order accurate BDF1 (AKA Backward Euler) method.
We choose h=0.5 and calculate the value of $q^{BDF1}(\Delta s;\Delta s)$ and $q^{BDF1}(\Delta s;0.5 \Delta s)$ with the `BDF1' subscript indicating that the values are calculated using the first order (n=1) BDF1 method.
Then we use Eqn.~\ref{eq:re} to calculate the \textit{2nd order accurate} value at $s=\Delta s$:
\begin{align}\label{eq:bdf2-re}
  q^{BDF1,RE}(\Delta s;\Delta s) =
  2*q^{BDF1}(\Delta s;0.5 \Delta s) -
  q^{BDF1}(\Delta s;\Delta s)
\end{align}
In the above equations, the superscript `$BDF1,RE$' denotes that the value is from BDF1 with Richardson Extrapolation performed once, yielding an extra order of accuracy.
This value is now used as the first point in the propagator solution.
All future values are calculated using the BDF2 scheme given in the main text.
There is some overhead associated with this method.
The $q^{BDF1}(\Delta s;\Delta s)$ term requires a single q calculation at $s=\Delta s$ (which would be needed even without Richardson extrapolation).
The $q^{BDF1}(\Delta s;0.5 \Delta s)$ term requires two q calculations ($s=0.5\Delta s$ and $s=\Delta s$) which are not otherwise required.
So this method retains the 2nd order accuracy at the expense of two additional propagator solve per polymer block.

\subsection{Backward Difference 3}
The BDF3 scheme is 3rd order accurate in contour and requires three previous contour points.
We present two options for using RE with BDF3.
First, we can follow to approach used for BDF2, except with use of the Crank-Nicolson scheme.
This requires only the first point to be determined using RE and following the same logic above (with n=2 in this case), gives the \textit{3rd order accurate} value at $s=\Delta s$:
\begin{equation}\label{eq:cn-re}
  q^{CN,RE}(\Delta s) =
  \frac{4*q^{CN}(\Delta s;0.5 \Delta s) - q^{CN}(\Delta s;\Delta s)}{3}
\end{equation}
Again this retains the higher order accuracy at the expense of two additional propagator solve per polymer block.

Alternatively, we can apply RE twice by starting with Eqn.~\ref{eq:bdf2-re} to get the 2nd order accurate values $q^{BDF1,RE}(\Delta s;\Delta s)$, $q^{BDF1,RE}(\Delta s;0.5\Delta s)$, $q^{BDF1,RE}(2\Delta s;\Delta s)$, and $q^{BDF1,RE}(2\Delta s;0.5\Delta s)$.
With those values, Eqn.~\ref{eq:re} (with n=2) is used to obtain 3rd order accurate values at $s=\Delta s$ and $s=2\Delta s$.
The equations are given in Eqns.~\ref{eq:re-bdf3-first} to \ref{eq:re-bdf3-last}.
The superscript `$BDF1,RE^2$' denotes that the value is from BDF1 with Richardson Extrapolation performed twice, yielding two extra orders of accuracy.
This approach requires calculating $q(s=2\Delta s)$ for three contour steps: $q(s=0.25\Delta s)$, $q(s=0.5\Delta s)$, $q(s=\Delta s)$.
The first requires eight q calculations, the second requires four, and the last requires the two q calculations (both of these would be preformed even if the extrapolation was not performed).
This leads to a total of 12 extra values per block in order to get the full 3rd order accuracy.
This is the same accuracy that came from Eqn.~\ref{eq:cn-re}, but it requires six times as many additional evaluations

\subsection{Backward Difference 4}
The BDF4 scheme is 4th order accurate in contour and requires four previous contour points\cite{cochran2006}.
As with BDF3, there are multiple ways to approach using Richardson extrapolation.
The approach we present uses the Crank-Nicolson values and the Richardson extrapolated Crank-Nicolson values (Eqn.~\ref{eq:cn-re}).
This approach requires calculating q values out to $s=3\Delta s$ for contour steps of $0.25\Delta s$, $0.5\Delta s$, and $\Delta s$.
This requires 18 extra q calculations per polymer block.
The equations for this approach are given in Eqns.~\ref{eq:re-bdf4-first} to \ref{eq:re-bdf4-last}

BDF3 equations with Richardson extrapolation
\begin{equation}
  q^{BDF1,RE}(\Delta s;0.5\Delta s) = 2*q^{BDF1}(\Delta s;0.25 \Delta s) -
  q^{BDF1}(\Delta s;0.5\Delta s) \label{eq:re-bdf3-first}
\end{equation}
\begin{equation}
  q^{BDF1,RE}(\Delta s;\Delta s) = 2*q^{BDF1}(\Delta s;0.5 \Delta s) -
  q^{BDF1}(\Delta s;\Delta s)
\end{equation}
\begin{equation}
  q^{BDF1,RE}(2\Delta s;0.5\Delta s) = 2*q^{BDF1}(2\Delta s;0.25\Delta s) -
  q^{BDF1}(2\Delta s;0.5\Delta s)
\end{equation}
\begin{equation}
  q^{BDF1,RE}(2\Delta s;\Delta s) = 2*q^{BDF1}(2\Delta s;0.5\Delta s) -
  q^{BDF1}(2\Delta s;\Delta s)
\end{equation}
\begin{equation}
  q^{BDF1,RE^2}(\Delta s) =
  \frac{4*q^{BDF1,RE}(\Delta s;0.5 \Delta s) - q^{BDF1,RE}(\Delta s;\Delta s)}{3}
\end{equation}
\begin{equation}
  q^{BDF1,RE^2}(2\Delta s) =
  \frac{4*q^{BDF1,RE}(2 \Delta s;0.5 \Delta s) -
  q^{BDF1,RE}(2 \Delta s;\Delta s)}{3} \label{eq:re-bdf3-last}
\end{equation}
BDF4 equations with Richardson extrapolation
\begin{equation}
  q^{CN,RE}(\Delta s;0.5\Delta s) =
  \frac{4*q^{CN}(\Delta s;0.25 \Delta s) - q^{CN}(\Delta s;0.5\Delta s)}{3} \label{eq:re-bdf4-first}
\end{equation}
\begin{equation}
  q^{CN,RE}(\Delta s;\Delta s) =
  \frac{4*q^{CN}(\Delta s;0.5 \Delta s) - q^{CN}(\Delta s;\Delta s)}{3}
\end{equation}
\begin{equation}
  q^{CN,RE}(2\Delta s;0.5\Delta s) =
  \frac{4*q^{CN}(2\Delta s;0.25\Delta s) - q^{CN}(2\Delta s;0.5\Delta s)}{3}
\end{equation}
\begin{equation}
  q^{CN,RE}(2\Delta s;\Delta s) =
  \frac{4*q^{CN}(2\Delta s;0.5\Delta s) - q^{CN}(2\Delta s;\Delta s)}{3}
\end{equation}
\begin{equation}
  q^{CN,RE}(3\Delta s;0.5\Delta s) =
  \frac{4*q^{CN}(3\Delta s;0.25\Delta s) - q^{CN}(3\Delta s;0.5\Delta s)}{3}
\end{equation}
\begin{equation}
  q^{CN,RE}(3\Delta s;\Delta s) =
  \frac{4*q^{CN}(3\Delta s;0.5\Delta s) - q^{CN}(3\Delta s;\Delta s)}{3}
\end{equation}
\begin{equation}
  q^{CN,RE^2}(\Delta s) =
  \frac{8*q^{CN,RE}(\Delta s;0.5 \Delta s) - q^{CN,RE}(\Delta s;\Delta s)}{7}
\end{equation}
\begin{equation}
  q^{CN,RE^2}(2\Delta s) =
  \frac{8*q^{CN,RE}(2 \Delta s;0.5 \Delta s) - q^{CN,RE}(2 \Delta s;\Delta s)}{7}
\end{equation}
\begin{equation}
  q^{CN,RE^2}(3\Delta s) =
  \frac{8*q^{CN,RE}(3 \Delta s;0.5 \Delta s) - q^{CN,RE}(3 \Delta s;\Delta s)}{7} \label{eq:re-bdf4-last}
\end{equation}
The final three equations give the 4th order accurate values for the first three contour points.
The alternative method using BDF1 values is not presented here, but it requires 42 extra q evaluations instead of the 18 required starting with the Crank-Nicolson values.

\subsection{Results with Richardson Extrapolation}
\begin{figure}
  \centering
  \includegraphics{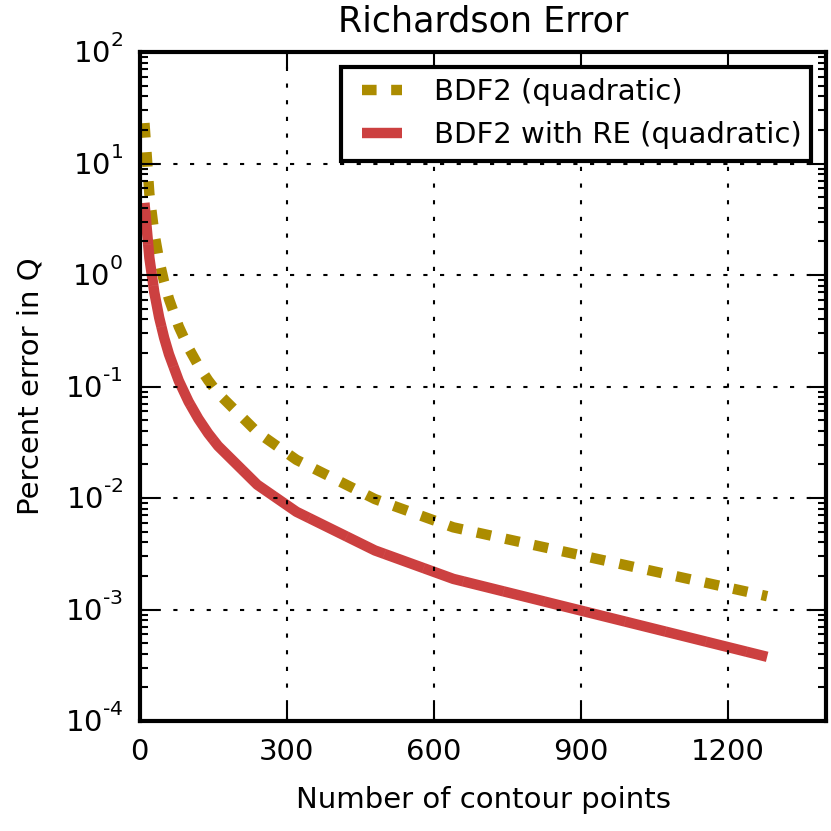}
  \caption{\label{fig:q_with_richardson} Error in partition function resulting from using initial points with lower order accuracy.
           Single unit cell of gyroid phase at f=0.4 with $32^3$ elements and quadratic basis functions.}
\end{figure}

The benefit of this extra effort to ensure the initial points have the desired accuracy is a lower error for a given number of contour points.
As before, we take the Q value of the spectral solution to be the true solution and using the error in Q as a measure of the solution accuracy.
While it may seem that a single point, out of potentially hundreds, being slightly off may not make a large difference, the error is substantial.
Figure \ref{fig:q_with_richardson} shows the error in Q with increasing contour points for the BDF2 method both with and without Richardson Extrapolation.
The rate of convergence to the exact solution is similar in both cases.
This is reasonable given that the order accuracy is the same for each curve except for the first value.
However, the error is significantly higher without RE applied.
It requires approximately twice as many points without RE to reach the given accuracy as it does with the extrapolation.
Given that there are only two extra contour points required to apply RE to the BDF2 scheme, it is clear that RE is worth the computational effort and complexity.

\bibliography{refs/scft}

\end{document}